\documentclass[fleqn,twoside]{article}
\usepackage{espcrc2}

\usepackage{graphicx}
\usepackage[figuresright]{rotating}

\newcommand{\AmS}{{\protect\the\textfont2
  A\kern-.1667em\lower.5ex\hbox{M}\kern-.125emS}}

\def\la{\langle}
\def\ra{\rangle}

\def\lb{\lbrack}
\def\rb{\rbrack}

 \def\Slash#1{
  \begin{picture}(5,6)(0,0)
  \put(-.7,-1.2){\line(5,6)6}
  \end{picture}
  \kern-.8em#1}
 \def\slash#1{
  \begin{picture}(5,6)(0,0)
  \put(-1.5,-1.7){\line(5,6)5}
  \end{picture}
  \kern-.8em#1}

\def\Sn{\Slash \nabla}

\def\gg5{\gamma_5}
\def\hg5{\hat{\gamma}_5}
\def\g4{\gamma_4}

\def\O{{\cal O}}

\def\U{{\cal U}}

\def\V{{\cal V}}

\def\Qlatmr1{Q_{lat}^{(m=r=1)}}

\def\be{\begin{eqnarray}}
\def\ee{\end{eqnarray}}

\def\bx{{\bf x}}
\def\bp{{\bf p}}
\def\t{\tau}
\def\hPsi{\hat{\Psi}}
\def\hU{\hat{U}}
\def\hD{\hat{D}}

\def\hU{\hat{U}}
\def\hcU{\hat{{\cal U}}}

\def\hV{\hat{{\cal V}}}
\def\hL{\hat{L}}

\title{Testing universality and the fractional power prescription for the staggered
fermion determinant}

\author{David H. Adams\address{Instituut-Lorentz for Theoretical Physics, 
        Leiden University, \\ 
        Niels Bohrweg 2, NL-2333 CA Leiden, The Netherlands}
\thanks{Supported by the European Commission, contract HPMF-CT-2002-01716.}
}
       
\begin{document}

\begin{abstract}
In Ref.~\cite{DA} expressions for the continuous Euclidean 
time limits of various lattice fermion determinants were derived and compared in order to test
universality expectations in Lattice QCD. Here we review that work with emphasis on its relevance
for assessing the fractional power prescription for the determinant in dynamical staggered 
fermion simulations. Some new supplementary material is presented; in particular the status of 
the ``universality anomaly'' in the determinant relations is clarified: it is shown to be
gauge field-independent and therefore physically inconsequential.  
\end{abstract}

\maketitle

\section{Introduction and background}

During the last decade there have been major developments in lattice gauge theory.
At the conceptual level, the long-standing problem of how to formulate chiral
symmetry on the lattice has been solved. The explicit solution is provided by the Overlap
fermion formulation \cite{ov}, which led to the Overlap Dirac operator for LQCD \cite{Neu1-2}
-- this has an exact chiral symmetry encapsulated in the Ginsparg--Wilson relation
\cite{GW} $\gamma_5D+D\gamma_5=aD\gamma_5D$, and is free \cite{laliena-Has(NPB)-Luscher(PLB)} 
of the problems caused by lack of chiral symmetry in previous formulations.
Implementing Overlap fermions in
numerical simulations is a very difficult challenge though, and it seems that substantial
advances in algorithms and computer power will be needed before realistic simulations with
dynamical Overlap fermions are possible.

In the meantime work continues on traditional fermion formulations.
A major practical development has been algorithm advances for improved staggered fermions 
which have made possible dynamical fermion simulations on realistically large lattices. 
An impressive increase in agreement with experimental values for various hadronic parameters
has been obtained compared to previous quenched simulations \cite{Davies(PRL)} 
(see, e.g., \cite{Davies-PhysToday,DeGrand}
for reviews and \cite{Bernard} for latest results).
However, as discussed previously by various authors \cite{Jansen,DeGrand,Neu04},
there are problematic conceptual/theoretical issues for dynamical staggered fermions
which raise the question of whether this is a first-principles approach
to QCD or whether it should instead be regarded as some kind of effective model.

The staggered formulation is a lattice theory for 4 degenerate continuum fermion flavours, and to represent
the fermion determinant of a single dynamical (sea) quark flavour one takes the fourth root
of the staggered fermion determinant. This prescription is based on the expectation that in
the continuum limit the staggered fermion determinant should factor into the product of 4 copies
of the fermion determinant for a single quark flavour. There are two theoretical issues here:
firstly, whether the expected factorisation actually occurs, and secondly, assuming it does occur,
whether the prescription can be fitted into the framework of local quantum field theory
at finite lattice spacing, i.e.
whether there exists a (exponentially-)local lattice Dirac operator $D$ such that
$det(D_{staggered})^{1/4}=detD$. 
Here we will focus on the factorisation issue. For discussion of the locality issue see
\cite{Jansen,DeGrand,Neu04} along with \S IX-D.7 of \cite{Bernard} and the references mentioned there.

It is well-known that terms in lattice actions which formally vanish for $a\to0$ can have
residual effects which remain in the continuum limit. Familiar examples include non-vanishing 
contributions to Feynman diagrams from ``irrelevant'' interaction terms in lattice perturbation theory
(see, e.g., \cite{Cap}) and the essential role of the Wilson term in reproducing the 
axial anomaly in LQCD with Wilson fermions \cite{Smit-Seiler}. In light of this, 
factorisation of the staggered fermion determinant in the continuum limit cannot be taken for 
granted -- it could conceivably be spoiled by residual effects from flavour-changing interactions.
It is therefore important to investigate, to the extent that it is possible and in as direct a way as
possible, whether the factorisation actually occurs.

Recall that the expansion of the logarithm of the fermion
determinant in powers of the gauge field can be expressed in terms of the one (fermion) loop gluonic
$n$-point functions. Therefore, the factorisation issue for the staggered fermion determinant is 
intimately tied to the question of whether perturbative LQCD with staggered fermions can be renormalised 
in a way that is consistent with 4 flavour QCD. The studies of perturbative LQCD with staggered fermions 
carried out to date indicate that this is the case,\footnote{Early indications of this came from 
evaluations of the fermion loop correction to the gluonic propagator \cite{Thun}, and the fermion 
self-energy \cite{Golterman-Gockeler}; see \cite{Cap} for a summary of subsequent work.}
and this can be taken as indirect evidence for factorisation of the staggered fermion determinant in
the continuum limit (although it says nothing about possible non-perturbative problems
for the factorisation).
It should be noted however that so far renormalisation 
of perturbative LQCD with staggered fermions to all orders in the gauge coupling has not been
demonstrated. (Some arguments
were sketched in \cite{Thun} but these are not a rigorous proof.) This is in contrast to the 
situation for Wilson fermions where renormalisability was demonstrated some time ago by Reisz
\cite{Reisz(NPB)} based on his power-counting theorem \cite{Reisz(CMP)}. As mentioned in \cite{Jansen},
the power-counting theorem, as it stands, does not apply to staggered fermions. 
This reflects the additional complications due to flavour-changing interactions in staggered fermion
perturbation theory. In light of all this, investigation of the factorisation issue for the staggered 
fermion determinant is also of interest as an indirect check on the possibility of renormalising
LQCD with staggered fermions to all orders in a way that is consistent with 4 flavour QCD.

The factorisation issue for the staggered fermion determinant is closely tied to a more general
universality issue: Is LQCD with a staggered fermion in the same universality class as, say,
LQCD with 4 flavours of Wilson fermions? This is certainly expected to be the case but should
nevertheless be tested wherever possible. One way to try to get insight into this, while at the
same time investigating the factorisation issue, would be to compare the staggered fermion
determinant with the fourth power of the Wilson fermion determinant in the continuum
limit. Ideally this should be done analytically (so as to be able to identify and discard physically
inconsequential factors which may be present in the determinants and which may either diverge
or vanish for $a\to0$); however this appears impossible with currently known techniques.
A {\em simplified} version of this test of universality is possible though \cite{DA} and we 
review it in the following. When supplemented with some additional results presented here, it allows 
us to analytically study the factorisation issue for a ``partially staggered'' fermion determinant
in a setting where only a partial continuum limit is required. From the results of
\cite{DA} (and the clarification of the ``universality anomaly'' which we give here) we find
that factorisation of the partially staggered fermion determinant does indeed occur in
this limit.

\section{A simplified framework for testing universality}

On a finite volume Euclidean spacetime lattice with temporal lattice spacing $a$ and spacial
lattice spacing $a'$ we consider the fermion action 
$S_{fermion}=a(a')^3\sum_{(\bx,\t)}\bar{\psi}(\bx,\t)D^{(r)}\psi(\bx,\t)$ 
where
\be
D^{(r)}&=&{\textstyle \frac{1}{a}}\gamma_4\nabla_4+{\textstyle \frac{r}{2a}}\Delta_4
+D_{space}+m \label{1a} \\
D_{space}&=&{\textstyle \frac{1}{a'}}\Sn_{space}+{\textstyle \frac{r'}{2a'}}\Delta_{space}
\nonumber
\ee
with $\Sn_{space}\!=\!\sum_{\sigma=1,2,3}\gamma_{\sigma}\nabla_{\sigma}\;$, $\Delta_{space}\!=\!
\sum_{\sigma=1,2,3}\Delta_{\sigma}\;$,  
$\nabla_{\mu}\psi(x)\!=\!{\textstyle \frac{1}{2}}\lb U_{\mu}(x)\psi(x\!+\!b\hat{\mu})
\!-\!U_{\mu}(x\!-\!b\hat{\mu})\psi(x\!-\!b\hat{\mu})\rb\;$, $\Delta_{\mu}\psi(x)\!=\!
2\psi(x)\!-\!U_{\mu}(x)\psi(x\!+\!b\hat{\mu})
\!-\!U_{\mu}(x\!-\!b\hat{\mu})^{-1}\psi(x\!-\!b\hat{\mu})\;$, 
$x\!=\!(\bx,\t)$ and $b=a$ or $a'$ as appropriate.
For $r\!=\!r'\ne0$ this is the Wilson fermion action, while for $r\!=\!r'\!=\!0$ it is the naive action.
In the case $r\!=\!0$, $r'\ne0$ it is a partially naive---partially Wilson action, with fermion 
doubling on the Euclidean time axis due to the absence of the temporal part of the Wilson term.
Universality can therefore be tested by comparing this lattice theory with the theory for
two flavours of Wilson ($r\ne0$) fermions: according to universality these should coincide in the
continuous Euclidean time limit $a\to0$. We do not need to take the spacial continuum limit
since the spacial parts of the actions of the two theories are the same.
Thus we have a simplified framework for testing universality
which only requires taking the continuum limit of one of the spacetime coordinates 
(chosen here to be time). The key advantage is that, as we will see, this is accessible to direct 
analytic investigation.

To connect this universality test with the factorisation issue we point out that
the $r\!=\!0$ theory has a partially staggered interpretation as follows. Introduce the staggered
'flavour' fields
\be
\psi_1(\bx,\t)=\psi(\bx,2\t)\ ,\quad \psi_2(\bx,\t)=\psi(\bx,2\t+a)
\nonumber
\ee
living on the partially blocked lattice with temporal spacing $2a$. In the free field case
$D^{(r=0)}$ then takes the form
\be
D^{(0)}\Big({\textstyle {\psi_1 \atop \psi_2}}\Big)
=\Big\lb\,{\textstyle \frac{1}{a}}\,\gamma_4\Big({\textstyle {0 \atop \partial_4^+}\;
{\partial_4^- \atop 0}}\Big)+D_{space}+m\Big\rb\Big({\textstyle{\psi_1 \atop \psi_2}}\Big) 
\nonumber
\ee
where $\partial_4^+$ ($\partial_4^-$) denotes the forward (backward) time difference
operator. Now, making a basis transformation in spinor$\otimes$flavour space specified by
$\O=\Big({\gamma_5\gamma_4 \atop -\gamma_5\gamma_4}\;{1 \atop 1}\Big)$ we obtain
\be
D^{(r=0)}\;\to\;\O^{-1}D^{(r=0)}\O \qquad\qquad\qquad\qquad\qquad&&\nonumber \\
=\gamma_4\!\otimes\!\Big({\textstyle {1 \atop 0}\;{0 \atop 1}}\Big)\,{\textstyle \frac{1}{2a}}
\,\partial_4+ 
\gamma_5\!\otimes\!\Big({\textstyle {0 \atop 1}\;{-\!1 \atop\,0}}\Big)\,
{\textstyle \frac{1}{2(2a)}}\,\Delta_4 &&\nonumber \\
\qquad\quad\quad\ +D_{space}+m \qquad\qquad\qquad\qquad\qquad(2)&&\nonumber
\ee
where $\partial_4=\frac{1}{2}(\partial_4^++\partial_4^-)$. This clearly has the structure of a 
'partially staggered' lattice Dirac operator: the 'time part' has the same form as the usual
free field staggered Dirac operator in the flavour field representation \cite{Kluberg}. In particular,
$\Delta_4$ comes with a non-trivial flavour matrix, and in the gauged theory this gives rise to
flavour-changing interactions. Thus we have a simplified setting for investigating
the question of whether residual effects from flavour-changing interactions spoil the factorisation 
of the staggered fermion determinant in the continuum limit. In the present setting, factorisation
of the partially staggered fermion determinant in the continuous time limit corresponds to the
following \hfill\break

\noindent {\em Universality expectation}:
\be
\lim_{a\to0}\;detD^{(r=0)}=\Big(\,\lim_{a\to0}\;detD^{(r\ne0)}\Big)^2 
\qquad\qquad(3)
\nonumber
\ee
(up to physically inconsequential factors (p.i.f.'s)). \hfill\break
This is something which can be checked analytically \cite{DA} as we describe
in the following sections.

\section{Results}

Set $\beta:=$length of the time axis (which we hold fixed when taking the 
continuous time limit $a\to0$); $\beta\!=\!aN_{\beta}$ where $N_{\beta}$
is the number of sites along the time axis; set $N:=$dimension of the quantum-mechanical Hilbert
space of spinor fields $\{\psi(\bx)\}$ living only on the spacial lattice, and take 
$U_4(\bx,\t)$ to be the lattice transcript of the 4-component $A_4(\bx,\t)$ of a smooth 
continuum gauge field. In this setup the following was shown in \cite{DA} 
(full details in \cite{DA(prep)}):

\noindent {\em Result of direct calculation}:
\be
detD^{(1)}\,\stackrel{a\to0}{\longrightarrow}\,
\big({\textstyle \frac{1}{a}}\big)^{NN_{\beta}}
e^{\frac{1}{2}\int_0^{\beta}TrM(\t)d\t}\,det({\bf 1}\!-\!\V(\beta)) &&\nonumber \\
(4) &&\nonumber\\
detD^{(0)}\,\stackrel{a\to0}{\longrightarrow}\,\big({\textstyle \frac{1}{2a}}\big)^{NN_{\beta}}
\,det({\bf 1}-\V(\beta))^2 \quad\ \ \qquad(5)&&\nonumber
\ee 
with the ingredients defined below. The gauge fields are assumed to satisfy periodic time boundary 
condition, and for simplicity we have stated the results (4)--(5) for the case where the fermion 
fields are also time-periodic. (The treatment in \cite{DA} covers the case of general time b.c.
$\psi(\bx,\t)=e^{-\alpha\beta}\psi(\bx,0)$. For $\alpha\!=\!\mu\!+\!i\pi/\beta$ this
corresponds to QCD at finite temperature $1/\beta$ and chemical potential $\mu$.)
In (4),
\be
M(\t):={\textstyle \frac{r'}{2a'}}\Delta_{space}(\t)+m\ \qquad\qquad\qquad\qquad\ (6)
\nonumber
\ee
is a linear map on the QM Hilbert space $\{\psi(\bx)\}$,
with $\Delta_{space}(\t)\psi(\bx)$ given simply by replacing $\psi(\bx,\t)$ with $\psi(\bx)$ in the 
definition of $\Delta_{space}\psi(\bx,\t)$. The $\V(\beta)$ in (4)--(5) is also an operator
on the QM Hilbert space; it is the specialisation to $\t\!=\!\beta$ of the operator
$\V(\t)$ defined $\forall\t\in{\bf R}$ as follows. We regard $\psi(\bx,\t)$ as a function 
$\Psi(\t)$ taking values in the QM Hilbert space, and introduce the continuous time---lattice space
Dirac operator
\be
D=\gamma_4(\,{\textstyle \frac{\partial}{\partial\t}}+A_4(\t))+D_{space}(\t)+m
\qquad\ \ \quad(7)&& \nonumber 
\ee
After extending the gauge field from $\t\in[0,\beta]$ 
to all $\t\in{\bf R}$, periodic under $\t\to\t\!+\!\beta$, this 
operator acts on QM Hilbert space-valued functions $\Psi(\t)$ defined for all $\t\in{\bf R}$.
Solutions to $D\Psi(\t)\!=\!0$ ($\t\in{\bf R}$, no periodicity requirement on $\Psi(\t)$)
are then specified by an initial value $\Psi(0)$, and $\V(\t)$ is defined to be the evolution
operator which determines the full solution from the initial value, i.e.
$\Psi(\t)=\V(\t)\Psi(0)$. This completes the description of the ingredients in (4)--(5). 
We outline the derivations of (4)--(5) in the next section.

The results (4)--(5) allow us to check to what extent the universality expectation (3) is satisfied.
The appearance of $det({\bf 1}-\V(\beta))$ in (4) versus $det({\bf 1}-\V(\beta))^2$ in (5) is clearly
in accordance with this expectation. The divergent factors in (4)--(5) are also in accordance
with it: given that the divergent factor in (4) is $(1/a)^{NN_{\beta}}$, and in light of the partially
staggered interpretation of the $r\!=\!0$ theory, the divergent factor in (5) should be
$((1/\tilde{a})^{N\tilde{N}_{\beta}})^2$ where $\tilde{a}\!=\!2a$ and 
$\tilde{N}_{\beta}\!=\!N_{\beta}/2$. This is precisely the divergent factor 
$(1/2a)^{NN_{\beta}}$ appearing in (5). However, (4)--(5) reveal a potential breakdown of the 
universality expectation due to the factor $e^{\int_0^{\beta}Tr\,M(\t)\,d\t}$ appearing in (4)
which has no counterpart in (5). Thus it is important to clarify the physical significance, or
lack thereof, of this ``universality anomaly''. Recalling (6) we see that, up to a p.i.f.,
the anomaly factor is 
\be
\exp\{{{\textstyle \frac{1}{2}}\int_0^{\beta}Tr({\textstyle \frac{r'}{2a'}}\Delta_{space}(\t)) d\t}\}
\qquad\qquad\qquad\ (8)
\nonumber
\ee
This involves the spacial link variables and cannot therefore be immediately dismissed as a p.i.f.
One might argue that it becomes a p.i.f. when the spacial continuum limit $a'\to0$ is taken
since $\frac{1}{a'}\Delta_{space}$ formally vanishes in this limit. This is a delicate issue 
though, since $Tr(\frac{1}{a'}\Delta_{space})$ actually diverges in this limit (the largest
eigenvalue of $\frac{1}{a'}\Delta_{space}$ is $\sim\frac{1}{a'}$). In \cite{DA} the status of this
anomaly factor was left unresolved. We now clarify the situation, showing that the anomaly is
in fact physically inconsequential. This is done by showing that $Tr\Delta_{space}(\t)$, and hence the 
anomaly itself, are {\em independent of the gauge field}, i.e. do not depend on the spacial link
variables entering in $\Delta_{space}(\t)$. To see this, recall that 
$\Delta_{space}(\t)\!=\!\sum_{\sigma=1,2,3}\Delta_{\sigma}(\t)$ with
$\Delta_{\sigma}(\t)\psi(\bx)=2\psi(\bx)-U_{\sigma}(\bx,\t)\psi(\bx+a'\hat{\sigma})
-U_{\sigma}(\bx-a'\hat{\sigma})^{-1}\psi(\bx-a'\hat{\sigma})$.
Evaluating $Tr\Delta_{\sigma}(\t)$ in a spacial plane wave basis $\psi_{\bp}(\bx)\sim e^{i\bp\bx}$
we get (ignoring overall normalisation factors)
\be
Tr\Delta_{\sigma}(\t)=\sum_{\bp}\la\psi_{\bp}\,,\Delta_{\sigma}\psi_{\bp}\ra\,=
\qquad\qquad\qquad\qquad&& \nonumber \\
\sum_{\bp}\sum_{\bx}\lb\,2\!-\!U_{\sigma}(\bx,\t)e^{ip_{\sigma}a'}\!-\!
U_{\sigma}(\bx\!-\!a'\hat{\sigma})^{-1}e^{-ip_{\sigma}a'}\rb&&
\nonumber
\ee
After interchanging the sums over $\bp$ and $\bx$ the terms involving the link variables are seen to 
vanish since $\sum_{p_{\sigma}}e^{ip_{\sigma}a'}=\delta(a')=0$.

Thus we have established that the universality expectation (3) does indeed hold up to
physically inconsequential factors, at  least when $r\!=\!1$ in the right-hand side of (3).
This is at the same time a demonstration of the factorisability
of the partially staggered fermion determinant in the partial continuum limit $a\to0$.
For technical reasons mentioned below we have only been able to evaluate the
$a\to0$ limit of $detD^{(r)}$ in the cases $r\!=\!0$, Eq.(5), and $r\!=\!1$, Eq.(4),  
but not yet for general $r\ne0$.

\section{Derivation of the results -- an outline}

The time-periodic lattice spinor fields $\psi(\bx,\t)$ on which $D^{(r)}$ acts are identified
with QM Hilbert space-valued functions $\Psi(\t)$ living on the temporal lattice sites, and these
can in turn be represented by vectors $\hPsi=(\hPsi(0),\dots,\hPsi(N_{\beta}-1))$ where
$\hPsi(k)\!:=\!\Psi(ak)$. Then $D^{(r)}$ is represented by 
\be
\hD^{(r)}\hPsi(k)=\qquad\qquad\qquad\qquad\qquad\qquad\qquad\qquad\nonumber\\
d_{-1}^{(r)}(k)\hPsi(k\!-\!1)+d_0^{(r)}(k)\hPsi(k)+
d_1^{(r)}(k)\hPsi(k\!+\!1)
\nonumber
\ee
where the $d_j^{(r)}(k)$'s are linear operators on the QM Hilbert space given by
$d_1^{(r)}(k)\!=\!\frac{1}{2a}(\gamma_4\!-\!r)\hU_4(k)\,$, $d_{-1}^{(r)}(k)\!=\!
-\frac{1}{2a}(\gamma_4\!+\!r)\hU_4(k\!-\!1)^{-1}\,$, $d_0^{(r)}(k)\!=\!\frac{r}{a}\!+\!
\hD_{space}(k)\!+\!m$ with $\hU_4(k)\!:=\!U_4(ak)$ and $\hD_{space}(k)\!:=\!D_{space}(ak)$.
(Here $U_4(ak)\psi(\bx):=U_4(\bx,ak)\psi(\bx)$ etc.)
After writing $\hD^{(r)}$ as an $N_{\beta}\times N_{\beta}$ matrix, its determinant can be 
straightforwardly evaluated via the method of \cite{Gibbs}. The cases $r\!=\!\pm1$ and $r\ne\pm1$
require separate treatments due to the fact that $d_{\pm1}^{(r)}(k)$ is invertible when $r\ne\pm1$
but not when $r\!=\!\pm1$. For simplicity we restrict to $r\ge0$ (the $r\le0$ case is analogous),
then the results of the determinant calculations are (with $N_{\beta}$ even in the $r\ne1$ case)
\cite{DA,DA(prep)}:
\be 
detD^{(r\ne1)}=\qquad\qquad\qquad\ \ \qquad\qquad\qquad &&\nonumber\\
({\textstyle \frac{(1-r^2)^2}{2a}})^{NN_{\beta}}\,det\big(
\big({\textstyle {{\bf 1} \atop 0}\;{0 \atop {\bf 1}}}\big)-\hcU^{(r)}(N_{\beta}/2)\big)
\ &&(9)
\nonumber \\
detD^{(1)}=({\textstyle \frac{1}{a}})^{NN_{\beta}}\,\chi(M)\,
det({\bf 1}-\hV(N_{\beta})) \ &&(10)
\nonumber
\ee
where $\chi(M):=\prod_{k=0}^{N_{\beta}-1}det({\bf 1}+aM(ak))^{1/2}$ and $\hcU^{(r)}(n)$ and 
$\hV(k)$ are the lattice evolution operators in the $r\ne1$ and $r\!=\!1$ cases, respectively:
\be
\hD^{(r\ne1)}\hPsi(k)=0&\Leftrightarrow&
\big({\textstyle {\hPsi(2n) \atop \hPsi(2n+1)}}\big)=\hcU^{(r)}(n)
\big({\textstyle {\hPsi(0) \atop \hPsi(1)}}\big) \nonumber \\
\hD^{(1)}\hPsi(k)=0&\Leftrightarrow&\hPsi(k)=\hV(k)\hPsi(0)
\nonumber
\ee 
($k\in{\bf Z}$, no periodicity requirement on $\hPsi(k)$). Note that in the $r\ne1$ case two initial
values are needed to determine the solution, hence $\hcU^{(r)}(n)$ is a 2$\times$2 matrix whose 
entries are linear operators on the QM Hilbert space, whereas in the $r\!=\!1$ case it turns out 
that only one initial value is needed.

The next step is to show that the lattice time evolution operators converge to appropriate
continuous time evolution operators in the $a\to0$ limit. The idea is to write $\hD^{(r)}$ 
in the form
\be
\hD^{(r)}=\hL_1(k){\textstyle \frac{1}{a}}\partial+\hL_0(k)\qquad\qquad\qquad\qquad\quad(11)
\nonumber
\ee
or some suitable variant thereof. Here $\partial$ is the forward or backward finite difference operator 
and $\hL_j(k)\,$ ($j\!=\!1,2$) are linear operators on the QM Hilbert space, parameterised by
$k\in{\bf Z}$, periodic under $k\to k\!+\!N_{\beta}\,$, 
such that
\be
\hL_j(k)=L_j(ak)+O(a)\qquad\qquad\qquad\qquad\ \quad(12)
\nonumber
\ee
for some operators $L_j(\t)$ parameterised by continuous time variable $\t\in{\bf R}$ with
periodicity under $\t\to\t\!+\!\beta$.
Then the solutions to $\hD^{(r)}\hPsi(k)=0$ approximate the solutions to $D\Psi(\t)=0$
where 
\be
D=L_1(\t){\textstyle \frac{d}{d\t}}+L_0(\t)\qquad\qquad\qquad\qquad\qquad(13)
\nonumber
\ee
Consequently the lattice evolution operator $\hcU(k)$ for solutions to $\hL\hPsi\!=\!0$
approximates the continuous time evolution operator $\U(\t)$ for solutions to $D\Psi\!=\!0$. 
(Explicitly, $\U(\t)=Te^{-\int_0^{\t}L_1(t)^{-1}L_0(t)\,dt}$ where $T=t$-ordering.)
In particular one has the {\em convergence theorem}: $\lim_{a\to0}\,\hcU(N_{\beta})=\U(\beta)$. 
This type of convergence result is well-known in the mathematics literature \cite{Burg-Forman}.
It is the key to deriving the results (4)--(5) from the determinant formulae (9)--(10).

We have so far only been able to write $\hD^{(r)}$ in the form (11) in the $r\!=\!0$ and
$r\!=\!1$ cases. The explicit expressions in these cases are given below. The problem of evaluating
$\lim_{a\to0}\,detD^{(r)}$ for general values of $r$ remains for future work; new techniques beyond
those described here may be required for this.

\underline{$r=1$ case}:  Decomposing $\hPsi\!=\!\Big({\hPsi_+ \atop \hPsi_-}\Big)\,$,
$\gamma_4\hPsi_{\pm}=\pm\hPsi_{\pm}\,$, we have 
\be
\hD^{(1)}&=&\hL_1\,{\textstyle \frac{1}{a}}\,\Big({\textstyle
{\partial^- \atop 0}\;{0\atop\partial^+}}\Big)+\hL_0
\nonumber \\
\hL_1(k)&=&\gamma_4\Big({\textstyle {\hU_4(k-1)^{-1} \atop 0}\;{0\atop \hU_4(k)}}\Big)
\nonumber \\ 
\hL_0(k)&=&\gamma_4\Big({\textstyle {\frac{1}{a}(1-\hU_4(k-1)^{-1}) \atop 0}\;
{0\atop \frac{1}{a}(\hU_4(k)-1)}}\Big) \nonumber \\
&&+\hD_{space}(k)+m
\nonumber 
\ee
The $\hL_j(k)$'s are periodic under $k\to k\!+\!N_{\beta}$ and
satisfy (12) with $L_1(\t)\!=\!\gamma_4$ and 
$L_0(\t)=\gamma_4A_4(\t)+D_{\space}(\t)+m$,
hence the continuous time--lattice space operator (13) is precisely the operator $D$ introduced
earlier in (7). The convergence theorem then gives $\lim_{a\to0}\hV(N_{\beta})=\V(\beta)$. 
The claimed result (4) now follows from (10) after noting that
$\lim_{a\to0}\,\chi(M)=e^{\frac{1}{2}\int_0^{\beta}Tr\,M(\t)\,d\t}$ \cite{DA}.

\underline{r=0 case}: Introducing the staggering \\
$\left({\hPsi_1(n) \atop \hPsi_2(n)}\right)=\left({\hPsi(2n) \atop \hPsi(2n+1)}\right)$ we have
\be
\hD^{(r=0)}\;=\;\hL_1\,{\textstyle \frac{1}{2a}}\,\Big({\textstyle
{\partial^+ \atop 0}\;{0\atop\partial^-}}
\Big)+\hL_0\qquad\ \qquad\qquad\qquad&&
\nonumber \\
\hL_1(n)\;=\;\gamma_4\Big({\textstyle {0\atop \gamma_4\hU_4(2n+1)}\;
{\gamma_4\hU_4(2n-1)^{-1} \atop 0}}\Big)\qquad\qquad\quad&&
\nonumber \\
\hL_0(n)\;= \qquad\qquad\qquad\qquad\qquad\qquad\qquad\qquad\qquad&&\nonumber \\
\Big({\textstyle {\hD_{space}(2n)+m \atop \gamma_4\frac{1}{2a}\,(\hU_4(2n)-\hU_4(2n-1)^{-1})}
{\gamma_4\frac{1}{2a}\,(\hU_4(2n+1)-\hU_4(2n)^{-1}) \atop \hD_{space}(2n)+m}}\Big)&&
\nonumber
\ee
The $\hL_j(n)$'s in this case are periodic under $n\to n\!+\!N_{\beta}/2$. Furthermore, 
$\hL_j(n)=L_j(2an)+O(a)$ with the corresponding continuous time---lattice space operator (13) 
given by \cite{DA}
\be
\widetilde{D}=
\left(\,{D_{space}(\t)+m \atop \gamma_4(\frac{d}{d\t}+A_4(\t))}
\ \ {\gamma_4(\frac{d}{d\t}+A_4(\t)) \atop D_{space}(\t)+m}\,\right)
\nonumber 
\ee
This can be rewritten as $\widetilde{D}=\O^{-1}\Big({D \atop 0}\;{0 \atop D}\Big)\O$ 
with $D$ as in (7) and $\O$ as defined above Eq.(2). 
The convergence theorem can now be applied to get
$\lim_{\tilde{a}\to0}\,\hU^{(r=0)}(\tilde{N}_{\beta})=
\O\Big({\V(\beta) \atop 0}\;{0 \atop \V(\beta)}\Big)\O^{-1}$, where 
$\tilde{a}\!=\!2a$ and $\tilde{N}_{\beta}\!=\!N_{\beta}/2$.
This together with (9) gives the claimed result (5).

Detailed rigorous proofs of the convergence theorem in the specific cases where it has been used 
above are given in \cite{DA(prep)}.

\section{Free field case}

In the free field case the evolution operator $\V(\beta)$ has a simple expression:
$\V(\beta)=e^{-\beta H}$ where $H$ is the QM Hamiltonian, i.e. 
$H=\gamma_4D_{space}=\gamma_4(\Sn_{space}+M)\,$, $M\!=\!\frac{r'}{2a'}\Delta_{space}+m$.
The energy eigenvalues of $H$ are $\pm E(\bp)=\sqrt{\bp^2+M(\bp)}\;$ (2-fold degenerate), 
and it follows that
\be
det({\bf 1}-\V(\beta))&=&\prod_{\bp}\big\lb\,(1-e^{\beta E})(1-e^{-\beta E})\big\rb^2
\nonumber \\
&=&\prod_{\bp}\big\lb\,e^{\beta E}\,(1-e^{-\beta E})^2\big\rb^2
\nonumber
\ee
Substituting this into (4)--(5) we obtain expressions for the $a\to0$ limits of $detD^{(1)}$
and $detD^{(0)}$ in terms of the $E(\bp)$'s. These expressions can be re-derived starting
from $detD^{(r)}=\{\mbox{{\em product of eigenvalues}}\}$, using Matsubara frequency summation
techniques \cite{DA(prep)}. This provides a crosscheck on the correctness of the results.

\section{Ambiguous continuous time limit of the partially staggered fermion determinant}

In obtaining the formula (9) for $detD^{(r=0)}$ the number $N_{\beta}$ of lattice sites along
the time axis was assumed to be even, and it was under this restriction that the $a\to0$ limit 
result (5) was obtained. One might expect that the $a\to0$ limit of $detD^{(0)}$ with
odd $N_{\beta}$ is the same. However, this turns out not to be the case.
For {\em odd} $N_{\beta}$ one finds \cite{DA(prep)}, instead of (9),
\be
detD^{(r\ne1)}=\qquad\qquad\qquad\qquad\qquad\qquad\qquad\qquad &&\nonumber\\
({\textstyle \frac{(1-r^2)^2}{2a}})^{NN_{\beta}}\,det\big\lb
\big({\textstyle {1 \atop 0}\;{0 \atop 1}}\big)-
\big({\textstyle {0 \atop 1}\;{1 \atop 0}}\big)
\,\hcU^{(r)}(N_{\beta}/2)\big\rb&&
\nonumber
\ee
where the precise meaning of $\hcU^{(r)}(N_{\beta}/2)$ in the odd $N_{\beta}$ case is given
in \cite{DA(prep)} and we simply mention here that $\hcU^{(0)}(N_{\beta}/2)$ 
has the same $a\to0$ limit as in the even $N_{\beta}$ case. It then follows by a straightforward 
calculation \cite{DA(prep)} that in this case
\be
detD^{(0)}\,\stackrel{a\to0}{\longrightarrow}\,\big({\textstyle \frac{1}{2a}}\big)^{NN_{\beta}}
\,det({\bf 1}\!-\!\V(\beta))\,det({\bf 1}\!+\!\V(\beta))
\nonumber
\ee
Comparing with (5) we see that
one of the factors $det({\bf 1}-\V(\beta))$ in (5) is replaced here by $det({\bf 1}+\V(\beta))$.
Thus the continuous time limit of the partially staggered fermion determinant 
$detD^{(0)}$ is ambiguous: different answers are obtained depending on whether $N_{\beta}$
is restricted to be even or odd.

This ambiguity has a natural explanation though: it reflects the fact that even $N_{\beta}$ and
odd $N_{\beta}$ in the $r\!=\!0$ lattice theory correspond to different  
time boundary conditions in the underlying continuous time theory. 
This is seen from the partially staggered `flavour field' interpretation as follows.
For even $N_{\beta}$ the flavour fields are time-periodic just like the original field, but for
odd $N_{\beta}$ we have
\be
\hPsi_1(0)=\hPsi(0)=\hPsi(N_{\beta})=\hPsi_2((N_{\beta}-1)/2) \ \ \qquad && \nonumber \\ 
\hPsi_2(0)=\hPsi(1)=\hPsi(N_{\beta}+1)=\hPsi_1((N_{\beta}+1)/2) \quad && \nonumber
\ee
I.e. $\Big({\Psi_1(0) \atop \Psi_2(0)}\Big)=\Big({\Psi_2(a(N_{\beta}-1)/2) \atop 
\Psi_1(a(N_{\beta}+1)/2)}\Big)$, which in the $a\to0$ limit becomes
$\Big({\Psi_1(\beta/2) \atop \Psi_2(\beta/2)}\Big)=\Big({0 \atop 1}\;{1 \atop 0}\Big)
\Big({\Psi_1(0) \atop \Psi_2(0)}\Big)$. Thus the b.c.'s in the underlying continuous time
theory are ``twisted'' in this case.

\section{Conclusions}

We have reviewed a test of universality where a ``partially naive'' lattice fermion theory with
fermion doubling on the Euclidean time axis is compared to the theory for two degenerate flavours
of Wilson fermions \cite{DA}. The continuous Euclidean time limits of the fermion determinants were
evaluated and compared (with Wilson parameter $r\!=\!1$ for the Wilson fermion). From the expressions
obtained, the partially naive fermion determinant was seen to coincide with the 2 flavour Wilson
fermion determinant (i.e. the square of the single flavour Wilson fermion determinant) up to physically
inconsequential factors, thus confirming the universality expectation. 
In connection with this the ``universality anomaly'' found in \cite{DA} was shown to be physically
inconsequential.

Furthermore, the partially naive
theory was shown here to have a ``partially staggered'' interpretation, thus our result can be interpreted as 
demonstrating factorisation of a partially staggered fermion determinant on the partial continuum limit
$a\to0$. This is a first piece of analytic evidence suggesting that factorisation of the complete staggered 
fermion determinant occurs in the full continuum limit (a problem with factorisation in the
complete case could very well also show up as a problem in the partially staggered case).
In light of the discussion in \S1, our result can also be interpreted as providing a hint that
renormalisation of perturbative LQCD with (at least partially) staggered fermions to all orders,
consistent with 4 flavour QCD (or 2 flavour QCD in our partially staggered case), 
is possible.\footnote{We remark that renormalisation of a variant of partially staggered fermions
in the context of a chiral sigma model (an effective theory; no gauge fields present) has been
carried out in \cite{Pernici}.}

I would like to thank the organizers of Lattice 2004 for a very enjoyable and stimulating conference,
and for the opportunity to present this work in the plenary session. I also thank the NCTS at Taiwan
National University for hospitality and support during visits where some of this work was done,
and Pierre van Baal for feedback on the manuscript.

\end{document}